# WIND MEASUREMENTS IN MARS' MIDDLE ATMOSPHERE AT EQUINOX AND SOLSTICE: IRAM PLATEAU DE BURE INTERFEROMETRIC CO OBSERVATIONS

R. Moreno[1], E. Lellouch[1], T. Encrenaz[1], F.Forget[2], E.Chassefiere[2], F.Hourdin[2], S. Guilloteau[3]
1 LESIA, Observatoire de Paris; 5, place Jules Janssen 92195 Meudon, France
2 LMD, Université Jussieu PARIS VI
3 Observatoire de Bordeaux

**Introduction:**

Characterizing the Martian atmosphere is an essential objective to understand its meteorology and its climate. The lower atmosphere (< 40 km) and middle atmosphere (40-80 km) of Mars appear dynamically coupled at much higher levels than in the case of the Earth. The vertical extension of the weather phenomena is considerable with for example Hadley's cells reaching the top of the neutral atmosphere (120 km). The circulation in the middle atmosphere modifies the meteorology of the lower atmosphere, affecting the transport and climatic processes

Observations of the CO rotational lines at millimeter (mm) wavelengths (Clancy *et al* 1990) have strongly contributed in the study of the vertical distribution of this compound and the thermal profile in the atmosphere of Mars over 0-70 km. Single-dish observations of the CO Doppler line-shift have allowed direct wind measurements in the martian middle atmosphere near 50 km altitude (Lellouch *et al* 1991), but at a low spatial resolution (12'') enabling only an essentially hemispheric resolution of the martian disk. The use of mm interferometry has allowed us to better spatially resolve the Martian disk, in order to obtain wind maps of the middle atmosphere (Moreno *et al* 2001).

**Observations:**
During the last oppositions of Mars (1999, 2001 and 2003), we observed the planet with the IRAM Plateau de Bure Interferometer (PdBI). Observations were obtained in the CO(2-1) and CO(1-0) rotational lines at 230.538 and 115.271 GHz respectively, with a spectral resolution up to 40 kHz. Depending on the year, Mars' average angular size was between 9.5'' and 23'' and the spatial resolution in our CO(1-0) maps permit to resolve Mars in about 4x4 independent point (Fig. 1). The observations sampled various seasons (Ls=140, 196, 262 and 317), and different dust situations (clear, global storm, regional storm). The high spectral resolution allowed us to determine and map Doppler shifts (Fig. 2) -especially in the (1-0) line where the signal-to-noise is much higher than in the CO(2-1) line - and therefore to directly investigate Mars' middle atmosphere (at 57±12 km) circulation (Fig. 3).

We present here a review of these wind measurements obtained at the PdBI, as well as those from the IRAM 30m telescope (Encrenaz *et al*, this conference), and a comparison with a Mars General Circulation Model (Forget *et al* 1999). **The major result is that for all observed periods, we detect strong retrograde winds with a typical 100 m/s velocity in the western Equatorial region.**

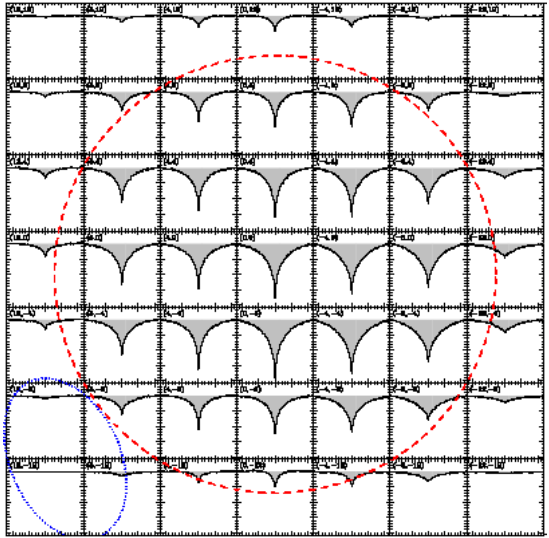

**Fig. 1** CO(1-0) spectral map observed on September 2003

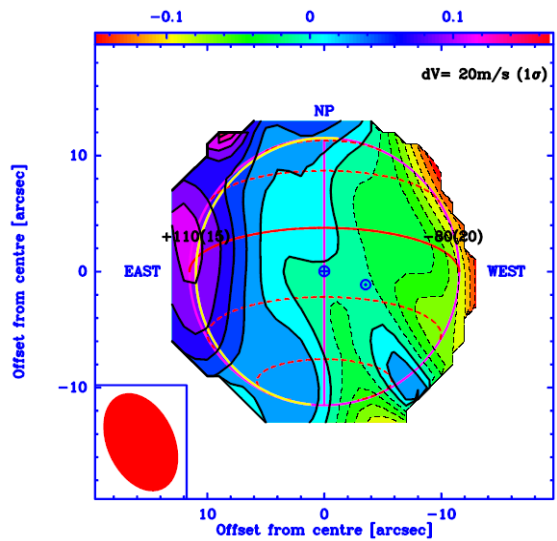

**Fig. 2** Line-of-sight winds derived from data presented in Figure 1. Strong retrograde winds are apparent with typical velocities of 100 m/s in the Equatorial region.

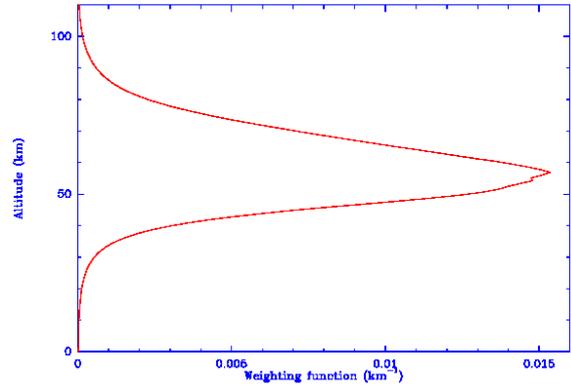

**Fig. 3** Weighting function of the Doppler CO(1-0) at the Martian limb with a beam size equal to 1/4 of the martian disk